\documentclass[aps,preprint,epsfig,rotate]{revtex4}
\usepackage{graphicx}
\usepackage{bm}
\usepackage{epsfig}

\input epsf
\begin{document}
\title{On the weakly-bound (1,1)-states in the three-body $d d \mu$ and $d t \mu$ muonic ions}

\author{Alexei M. Frolov}
 \email[E--mail address: ]{alex1975frol@gmail.com} 


\affiliation{Department of Applied Mathematics \\
 University of Western Ontario, London, Ontario N6H 5B7, Canada}

\date{\today}

\begin{abstract}

The both total and binding energies of the (1,1)-states in the weakly-bound three-body muonic $d d \mu$ and $d t \mu$ ions are determined to high numerical accuracy. The 
binding energy of the (1,1)-state in the muonic $d t \mu$ ion is evaluated as $\varepsilon(d t \mu)$ = -0.66033003831(30) $eV$, while for the same state in the muonic 
$d d \mu$ ion we have found that $\varepsilon(d d \mu)$ = -1.9749806166970(30) $eV$. These energies are the most accurate numerical values obtained for these systems and 
they are sufficient for all current and future experimental needs.  

\noindent 
PACS number(s): 36.10.Ee, 36.10.-k and 31.15.xt

\end{abstract}

\maketitle


\section{Introduction}

In this communication we report a number of new results of highly accurate computations of the weakly-bound (1,1)-states (or excited $P^{*}(L = 1)-$states) in the three-body 
$d d \mu$ and $d t \mu$ muonic ions. Some time ago these two weakly-bound states were of great interest for the development of the `resonance' approach to the muon-catalized 
nuclear fusion (see, e.g., \cite{MS}). On the other hand, these two states can be considered as ideal examples of very weakly-bound three-body Coulomb systems. The dimensionless 
ratio $\tau$ of the binding energies ($\varepsilon$) of these two states in the $d d \mu$ and $d t \mu$ ions to their total energies ($E$) is substantially less than 0.01 (or 
1 \%). The value of $\tau$ = 1 \% is the general criterion of weakly-boundness of any bound state. For the (1,1)-state in the $d t \mu$ muonic ion one finds $\tau \le$ 0.00245 
(or $\tau \le$ 0.245 \%). Highly accurate and precise computations of such bound states is a very difficult task, which, however, is of great interest by itself as well as in 
numerous applications. Also, our results of highly accurate computations of the weakly-bound (1,1)-states in the three-body $d d \mu$ and $d t \mu$ muonic ions are of interest 
for the future development of the general theory of bound states in the Coulomb three-body systems with unit charges. The total and binding energies obtained in this study for 
the weakly-bound (1,1)-states in the three-body $d d \mu$ and $d t \mu$ muonic ions can be considered as the final results for these systems. All systems mentioned in this study 
are considered within the framework of the non-relativistic three-body Coulomb problem.

Our goal in this study is to evaluate the total energies of the weakly-bound (1,1)-states in the three-body $d d \mu$ and $d t \mu$ muonic ions. Here we designate these states by 
using the $(\ell,\nu)-$notation which are more appropriate for the two-center quasi-molecular systems. In this notation, the symbols $\nu$ and $\ell$ stand for the quantum numbers
`vibrational' and `rotational' quantum numbers, respectively. Rigorously speaking, these two quantum numbers ($\nu$ and $\ell$) correspond to the two-center, pure adiabatic 
three-body system, e.g., these quantum numbers are good for the three-body, one-electron ${}^{\infty}$H$_{2}^{+}$ ion which contains two infinite nuclear masses (or Coulomb 
centers). In actual muonic ions $d d \mu$ and $d t \mu$ these quantum numbers are not conserved. Nevertheless, the use of the `rotational' $\ell$ and `vibrational' $\nu$ quantum 
numbers allows one unambigously to designate an arbitrary bound state in all three-body muonic ions $a b \mu$, where $a$ and $b$ are the nuclei of hydrogen isotopes $p, d$ and/or 
$t$. The system of $(\nu, \ell)-$notations has a number of advantages in applications to the three-body muonic ions $a b \mu$. It can be seen, if we compare such a system of 
notations with the one-center (or atomic) system of notations, where the same weakly-bound (1,1)-states in the three-body $d d \mu$ and $d t \mu$ muonic ions are designated as the 
excited $P^{*}(L = 1)-$states. 

By using three different sets of masses of the bare nuclei of deuterium, tritium and negatively charged muon we determine the both total and binding energies of the weakly-bound 
(1,1)-states in the three-body $d d \mu$ and $d t \mu$ muonic ions to very high numerical accuracy. For the $d d \mu$ and $d t \mu$ ions the total ($E$) and binding ($\varepsilon$)
energies are related by the following relations: $\varepsilon_{\nu\ell}(d d \mu) =  (2 Ry) [E_{\nu\ell}(d d \mu) - E_{1 S}(d \mu)]$ (for the $d d \mu$ ion) and 
$\varepsilon_{\nu\ell}(d t \mu) = (2 Ry) [E_{\nu\ell}(d t \mu) - E_{1 S}(t \mu)]$ (for the $d t \mu$ ion), where $E_{1 S}$ is the total energy of the ground $1 S$-state of the 
heaviest muonic atom, i.e., $d \mu$, or $t \mu$ muonic-atom. All these total energies must be taken in muon-atomic units (or $m.a.u.$, where $\hbar = 1, e = 1$ and $m_{\mu} = 1$), 
while $2 Ry$ is the conversion factor from $m.a.u.$ to $eV$ (electron-volts), since usually all binding energies are expressed in electron-volts. 

\section{Variational method}

The non-relativistic Hamiltonian of the $d d \mu$ and $d t \mu$ muonic ions takes the form
\begin{eqnarray}
 H = -\frac{\hbar^2}{2 m_{\mu}} \Bigl[ \Bigl(\frac{m_{\mu}}{M_d}\Bigr) \nabla^2_1 + \Bigl(\frac{m_{\mu}}{M_{d;t}}\Bigr) \nabla^2_2 + \nabla^2_3 \Bigr] - \frac{e^2}{r_{32}} - 
 \frac{e^2}{r_{31}} + \frac{e^2}{r_{21}} \; \; \; , \; \; \label{Hamil3}
\end{eqnarray}
where $\hbar = \frac{h}{2 \pi}$ is the reduced Planck constant, $m_{\mu}$ is the mass of the negatively charged muon and $e$ is the absolute value of the electric charge of 
electron. In this equation the subscripts 1 and 2 designate the two nuclei of hydrogen isotopes. The index 1 stands for deuterium, while index 2 means deuterium ($M_{d;t} = M_d$) 
in the $d d \mu$ ion and tritium ($M_{d;t} = M_t$) in the $d t \mu$ ion. The subscript 3 denotes the negatively charged muon ($\mu^{-}$). The masses of these particles are 
designated as $m_{\mu}, M_{d}$ and $M_{t}$ (their numerical values are discussed below). For the both muonic ions $d d \mu$ and $d t \mu$ we need to solve the non-relativistic 
Schr\"{o}dinger equation $H \Psi = E \Psi$, where $H$ is the Hamiltonian, Eq.(\ref{Hamil3}), and $E (E < 0$) is the eigenvalue, which coincides with the total energy of the $d d 
\mu$ and/or $d t \mu$ ions in their weakly-bound (1,1)-states (or $P^{*}(L = 1)-$states). These states will be stable, if their total energy is lower than the total energy of the 
ground $1 S-$ state in the two-body $d \mu$ or $t \mu$ muonic atoms, respectively (see above). In muon-atomic units these energies equal $E_{d\mu}= -0.5 \Bigl(1 + 
\frac{m_{\mu}}{M_d}\Bigr)^{-1}$ (for the $d \mu$ atom) and $E_{t\mu}= -0.5 \Bigl(1 + \frac{m_{\mu}}{M_t}\Bigr)^{-1}$ (for the $t \mu$ atom). 

In our calculations of the (1,1)-states of the three-body $d d \mu$ and $d t \mu$ ions we apply our exponential variational expansion \cite{FroEf} (see also \cite{DK}) which was 
substantially modified to increase its overall accuracy and efficiency. Such modifications include the use of short-term cluster functions \cite{Fro98}, careful optimization of a 
large number of non-linear parameters of the method \cite{Fro2001}, additional optimization of the `fast' non-linear parameters and other similar steps. The explicit form of the 
exponential variational expansion in the perimetric/relative coordinates takes the form 
\begin{eqnarray}
 \Psi_{LM} &=& \frac{1}{\sqrt{2}} (1 + \delta_{21} \hat{P}_{21}) \sum_{i=1}^{N} \sum_{\ell_{1}} C_{i} {\cal Y}_{LM}^{\ell_{1},\ell_{2}}({\bf r}_{31}, {\bf r}_{32}) 
 \exp(-\alpha_{i} u_1 - \beta_{i} u_2 - \gamma_{i} u_3) \label{exp1} 
\end{eqnarray}
where $C_{i}$ are the linear (or variational) parameters, $\alpha_i, \beta_i, \gamma_i$ are the non-linear parameters. The operator $\hat{P}_{21}$ is the permutation operator of 
identical particles 1 and 2, e.g., the two deuterium nuclei in the $d d \mu$ ion. For the $d t \mu$ ion the presence of this operator in Eq.(\ref{exp1}) has no sense and we add the 
Kronecker delta $\delta_{21}$ to cancel the second term from the final expression. The notation ${\cal Y}_{LM}^{\ell_{1},\ell_{2}}({\bf r}_{31}, {\bf r}_{32})$ stands for the bipolar 
harmonics \cite{Varsh} which are explicitly defined in \cite{Varsh} (see also \cite{Fro2011}). Formally, without loss of generality, one can restrict to the consideration of the 
bipolar harmonics with $M = 0$. Then for $L = 1$ one finds the two families of bi-polar harmonics ${\cal Y}_{10}^{1,0}({\bf r}_{31}, {\bf r}_{32}) \simeq ({\bf k} \cdot {\bf r}_{21})$ 
and ${\cal Y}_{01}^{1,0}({\bf r}_{31}, {\bf r}_{31}) \simeq ({\bf k} \cdot {\bf r}_{32})$, where ${\bf r}_{ij} = {\bf r}_{i} - {\bf r}_{j}$ is the interparticle vector and ${\bf k}$ 
is the unit vector oriented along $z-$axis. Angular integration in all arising integrals is reduced in this case to the calculation of three `angular' integrals over directions of the 
${\bf k}-$vector (for more detail, see, e.g., \cite{Fro1987}). However, in actual numerical computations of any bound state with $L \ge 1$ it is better to apply the bipolar harmonics 
defined in \cite{Varsh}, since this method is more advanced, universal and reliable. The angular integrals of the products of bi-polar harmonics are always expressed by the scalar 
functions of the three relative coordinates $r_{32}, r_{31}, r_{21}$. This fact has a fundamental meaning for highly accurate computations of arbitrary three-body systems considered 
in the non-relativistic approximation (see discussion in \cite{Fro2015}). 

The relative coordinates $r_{ij}$, where $(ij) = (ji) = (32), (31), (21)$, are defined as follows $r_{ij} = \mid {\bf r}_{i} - {\bf r}_{j} \mid$, where ${\bf r}_{i}$ and ${\bf r}_{j}$ 
are the corresponding Cartesian coordinates of point particles and $i (\ne j)$ = (1, 2, 3). Also, in Eq.(\ref{exp1}) the notations $u_1, u_2, u_3$ stand for three perimetric 
coordinates which are simply related with the relative interparticle coordinates $r_{32}, r_{31}$ and $r_{21}$ by the following linear relations
\begin{eqnarray}
  & & u_1 = \frac12 ( r_{21} + r_{31} - r_{32}) \; \; \; , \; \; \; r_{32} = u_2 + u_3 \nonumber \\
  & & u_2 = \frac12 ( r_{21} + r_{32} - r_{31}) \; \; \; , \; \; \; r_{31} = u_1 + u_3 \; \; \; \label{coord} \\
  & & u_3 = \frac12 ( r_{31} + r_{32} - r_{21}) \; \; \; , \; \; \; r_{21} = u_1 + u_2 \nonumber
\end{eqnarray}
where $r_{ij} = r_{ji}$. In contrast with the relative coordinates $r_{32}, r_{31}, r_{21}$ the three perimetric coordinates $u_1, u_2, u_3$ coordinates (also called the Pekeris triangle 
coordinates \cite{Pek}) are independent of each other and each of them varies between 0 and $+\infty$. This drastically simplifies analytical and numerical computations of all three-body 
integrals which are needed for solution of the corresponding eigenvalue problem and for evaluation of a large number of bound state properties for different three-body systems. The main 
advantage of the perimetric coordinates follows from the fact that these three coordinates are the best (or `natural') coordinates to solve any problem in triangular geometry. The 
exponential variational expansion, Eq.(\ref{exp1}), is used to approximate the exact wave function $\Psi$ and solve the non-relativistic Schr\"{o}dinger equation $H \Psi = E \Psi$ (or 
minimize the energy functional $E = \min_{\Psi} (\frac{\langle \Psi \mid H \mid \Psi \rangle}{\langle \Psi \mid \Psi \rangle}$) for the Coulomb three-body system(s) with unit charges. 
The explicit form of the Hamiltonian $H$ can be found, e.g., in \cite{Fro98} and \cite{Fro2001}.  

\section{Numerical computations and results}

Results of our numerical computations of the total energies of the (1,1)-states in the three-body $d d \mu$ and $d t \mu$ muonic ions can be found in Table I. All computed energies are 
expressed in the muon-atomic units. In these calculations we have used the three following sets of particle masses. The old `set' of particle masses which are expressed in the electron 
mass $m_e$
\begin{eqnarray}
  m_{\mu} = 206.768262 \; m_e \; \; \; , \; \; \; m_p = 1836.152701 \; m_e \label{mass1} \\
  m_{d} = 3670.483014 \; m_e \; \; \; , \; \; \; m_t = 5496.92158 \; m_e \nonumber
\end{eqnarray}
and the `new' set of particle masses which are expressed in the high-energy mass units $MeV/c^{2}$. 
\begin{eqnarray}
 m_{\mu} = 105.65836668 \; \; \; , \; \; \; m_p = 938.272046 \label{mass2} \\
 m_{d} = 1875.612859 \; \; \; , \; \; \; m_t = 2808.920906 \nonumber
\end{eqnarray}
The particle masses from the old set were extensively used in calculations of the muonic three-body ions since the end of 1990's, while the masses from the new set were determined in 
high-energy experiments performed between 1998 and 2007. These `new' masses were also used in highly accurate computations of the bound states in muonic three-body ions (see, e.g., 
\cite{Fro2011}). The third set of particle masses contains evaluated values obtained in the most recent high-enenrgy experiments. These masses are currently recomended for scientific 
use by CODATA/NIST. The numerical values of masses from this `recent' set of particle masses are expressed in the high-energy mass units $MeV/c^{2}$
\begin{eqnarray}
 m_{\mu} = 105.6583745 \; \; \; , \; \; \; m_p = 938.2720813 \label{mass3} \\
 m_{d} = 1875.612928 \; \; \; , \; \; \; m_t = 2808.921112 \nonumber
\end{eqnarray}
These masses have never been used in earlier calculations of the muonic three-body ions $a b \mu$. Formally, we can assume that the particle masses from the third (or `recent') set 
provide the better numerical accuracy and overall quality of numerical results. To re-calculate the results (energies) from muon-atomic units to $eV$ we also need to apply the doubled 
Rydberg constant which equals $2 Ry$ = 27.211386018 $eV$ and the rest mass of electron $m_e$ = 0.510 998 9461 $MeV/c^{2}$ (the condition $2 Ry = \alpha^{2} m_e c^{2}$ = 
27.21138602876$\ldots$, where $\alpha$ is the fine-structure constant $\alpha$ = 7.297352568$\cdot 10^{-3}$, is always obeyed). 

Based on the results (total energies) from Table I we can predict the following numerical evaluations for the `exact' total energies $E$ of the weakly-bound (1,1)-states in the $d t \mu$ 
and $d d \mu$ ions (in muon-atomic units):
\begin{eqnarray}
  E_o(d t \mu) &=& -0.48199152997398(5) \; , \;  E_o(d d \mu) = -0.4736867338427636(5) \; \; \label{res1} \\
  E_n(d t \mu) &=& -0.48199152705475(5) \; , \;  E_n(d d \mu) = -0.4736867311211389(5)  \; \; \label{res2} \\
  E_r(d t \mu) &=& -0.48199152658965(5) \; , \;  E_r(d d \mu) = -0.4736867297731058(5)  \; \; \label{res3} 
\end{eqnarray} 
where all energies are presented in muon-atomic units, while the indeces $o$ and $n$ mean the `old', `new' and `recent' sets of particle masses. If we know the total energies $E$ of 
these two ions, then it is easy to determine the corresponding binding energies. In general, the binding energy of an arbitrary three-body muonic ion $a b \mu$ is the difference between 
its total energy $E(a b \mu)$ and the total energy of the heaviest muonic atom ($b \mu$) in its ground $S-$state. This means that we need to subtract the total energies of the ground 
$S-$state of the $t \mu$ and $d \mu$ atoms from the results shown in Eqs.(\ref{res1}) and (\ref{res2}), respectively (for more details, see discussion of binding energies of muonic 
molecules in \cite{FroEf} and \cite{Fro2011}). By calculating the corresponding binding energies for each of these systems and by applying the numerical value of double Rydberg mentioned 
above one finds from Eqs.(\ref{res1}) - (\ref{res3}) 
\begin{eqnarray}
  \varepsilon_o(d t \mu) &=& -0.66033844243(30) \; eV  \; , \; \varepsilon_o(d d \mu) = -1.9749873572476(30) \; eV \;  \label{binden1} \\
  \varepsilon_n(d t \mu) &=& -0.66033258710(30) \; eV  \; , \; \varepsilon_n(d d \mu) = -1.9749829079445(30) \; eV \; \label{binden2} \\
  \varepsilon_r(d t \mu) &=& -0.66033003831(30) \; eV  \; , \; \varepsilon_r(d d \mu) = -1.9749806166970(30) \; eV \; \label{binden3} 
\end{eqnarray} 
As follows from Eqs.(\ref{binden1}) - (\ref{binden3}) the binding energies computed for the weakly-bound (1,1)-state in the $d t \mu$ ion are $\approx$ 100 times less accurate than the 
analogous energies obtained for the the weakly-bound (1,1)-state in the $d d \mu$ ion. Nevertheless, these binding energies are substantially more accurate than the binding energies ever 
obtained for these weakly-bound (1,1)-states in the $d d \mu$ and $d t \mu$ ions, if the same particle masses were used in calculations (see, Eqs.(\ref{binden1}) and (\ref{binden2})). 
The `recent' set of particle masses have never been used in highly accurate computations of the three-body muonic ions.  

\section{Discussion and Conclusions}

The obtained values are the most accurate binding energies known for these systems in the literature (compare them with earlier values obtained in \cite{Fro2011}). In applications to the 
resonance muon-catalyzed $(d,t)-$fusion the following temperature $T_{dt\mu} = 1.16045221 \cdot 10^{4} \cdot \varepsilon(d t \mu)$ $K$ is of a great interest. By using our computed values 
one finds that $T_{dt\mu} \approx$ 7662.84410 $K$ (for the new set of masses, Eq.(\ref{binden2})). In the `resonance' muon-catalized nuclear fusion the energy released during formation of 
weakly bound (1,1)-state in the $d t \mu$ ion (for the $d d \mu$ ion analogously) is transfered directly into molecular excitations of the six-body molecular clusters such as $p (d t \mu) 
e_2, d (d t \mu) e_2$ and $t (d t \mu) e_2$. The main problem here is to avoid possible ionization, i.e., emission of the free electron from the arising quasi-molecular system. In general, 
formation of the six-body molecular cluster $d (d t \mu) e_2$ (or two-center quasi-molecule $d X e_2$, where $X = (d t \mu)$, for short) can proceed in the two following ways (for more 
details, see, e.g., \cite{MS}):  
\begin{eqnarray}
 t \mu + {\rm D}_2 &=& [d (d t \mu)] e_2 = \Bigl\{[d (d t \mu)] e\Bigr\}^{+} + e^{-} \; \; \; , \label{nonres} \\
 t \mu + {\rm D}_2 &=& \Bigl\{[d (d t \mu)] e_2\Bigr\}^{*}  \; \; \; \label{res}
\end{eqnarray}
where the notation $*$ means internal excitation of the quasi-molecular $[d (d t \mu)] e_2$ system, i.e., rotational and/or vibrational excitations of this two-center, six-body 
quasi-molecule. First reaction, Eq.(\ref{nonres}), proceeds very slow, since electron ionization during this process takes a long time. This (first) reaction corresponds to the 
non-resonance (or slow) process of the muon-catalyzed fusion \cite{Alv}. During such reactions the five-body $\Bigl\{[d (d t \mu)] e\Bigr\}^{+}$ ions are formed. The bound state properties 
of these five-body ions have never been investigated. 

The second process, Eq.(\ref{res}), does not lead to the emission of a free electron, i.e., in this case we cannot observe any ionization of the six-body quasi-molecule and formation of 
the five-body quasi-molecular $\{[d (d t \mu)] e\}^{+}$ ion. In general, ionization of the six-body quasi-molecule $[d (d t \mu)] e_2$ can be avoided, if the energy released during the 
formation of the three-body muonic $(d t \mu)^{+}$ ion is less than ionization energy of the six-body quasi-molecule. Furthermore, the value of released energy must be close (and even 
very close) to the excitation energy of the six-body quasi-molecular cluster $[d (d t \mu)] e_2$ mentioned above \cite{MS}. This means that numerical values from Eqs.(\ref{binden1}) and 
(\ref{binden2}) have to be approximately equal to the excitation energies of some vibrational and rotational energy levels in the two-center (quasi-adiabatic) molecular cluster 
$d (d t \mu) e_2$ (more details can be found in \cite{MS}). 

In other words, the resonance formation of the $d (d t \mu) e_2$ quasi-molecule will proceed in those cases, when this two-center system has a rotationally and vibrationally excited state 
with the temperature close (or very close) to the $T_{dt\mu}$ value. In actual experiments we deal with the instant `resonance' formation of different quasi-molecular six-body systems, 
since the $[ p (d t \mu)] e_2, [ d (d t \mu)] e_2$ and $[ t (d t \mu)] e_2$ systems are formed at the same time. Each of these six-body (two-center) quasi-molecules has slightly different 
vibrational and rotational energy levels (or levels with slightly different temperatures). This substantially complicates theoretical predictions and exprerimental measurments of the 
`exact' resonance temperature of the process. Here we cannot discuss this interesting problem in detail. Note only that the resonance muon-catalized fusion which includes formation of the 
$d t \mu$ ions in the weakly-bound (1,1)-state was observed in numerous experiments (see, e.g., \cite{Fus2}, \cite{Balin}, \cite{Breun} and references in \cite{MS}). The current (maximal) 
number of the muon-catalyzed nuclear reactions of $(d,t)-$fusion is evaluted as 155 $\pm$ 35 \cite{MS} per one muon involved in the process.   

Results obtained in this study can be considered as the final energies of the weakly-bound (1,1)-states in the three-body $d d \mu$ and $d t \mu$ ions. Further numerical improvement of 
these results is possible, but has no direct physical meaning. However, highly accurate computations of the basic geometrical and dynamical properties (as well as the lowest-order 
relativistic and QED corrections) for the weakly-bound (1,1)-states of these muonic ions $d d \mu$ and $d t \mu$ are truly needed in applications. On the other hand, direct comparison of 
our current results for the weakly-bound (1,1)-states in the three-body $d d \mu$ and $d t \mu$ ions with the results of earlier variational studies (see, e.g., \cite{FroEf}, \cite{BD1}, 
\cite{Kam} and references therein) illustrates an amazing progress which has been achieved in accurate bound state computations of three-body systems with arbitrary particle masses and 
electrical charges (non-atomic three- and few-body systems). Note also that by performing numerical computations for this study we have constructed the two very compact wave functions 
with small and relatively small number of terms in Eq.(\ref{exp1}). In particular, the trial wave functions, which contain $N$ = 400 and $N$ = 800 basis variational wave functions 
(exponents), respectively, provide the following total energies for the weakly-bound (1,1)-state in the $d t \mu$ ion: -0.4819915007954827 $m.a.u.$ ($N$ = 400) and -0.481991520513476 
$m.a.u.$ ($N$ = 800). These short-term cluster functions with the carefully optimized non-linear parameters have been used in our two-stage procedure to obtain results presented in 
Table I.  

Note also that during the actual muon-catalyzed nuclear fusion all three-body ions, including $(d d \mu)^{+}$ and $(d t \mu)^{+}$ ions, are formed and exist only as internal parts (or 
internal clusters) in the six-, five- and four-body quasi-molecular and quasi-atomic systems, e.g., $[d (d t \mu)] e_2, [d (d t \mu)] e, (d t \mu) e_2, (d t \mu) e$ and other similar 
systems. The energy spectrum and properties of the $(d d \mu)^{+}$ and $(d t \mu)^{+}$ muonic ions are noticeably changed due to direct electromagnetic interactions of these three-body 
ions with the electrons and bare nucleus of hydrogen isotopes which are included in the same four-, five- and six-body structure. This means that we need to study the internal structure 
of such five- and six-body systems and their bound state spectra \cite{FroA}. First attempts to evalute the bound state properties and total and/or binding energies of some five- and 
six-body systems, which include three-body muonic ions, have been made in \cite{FroA} (see also \cite{FroAA}), where we used the variational expansion written in multi-dimensional 
gaussoids of the relative inter-particle coordinates (this expansion was proposed and developed in \cite{KT}). By using some effective optimization strategies for the non-linear 
parameters in such variational expansions we could obtain a number of interesting results (for more details, see \cite{FroA}) for many bound states in the five-body (one-center) ions 
$a b \mu e_2$ and six-body (two-center) quasi-molecules $a b c \mu e_2$. Accurate numerical investigations of such five- and six-body systems open a new avenue in our understanding of 
many processes which occur during the muon-catalyzed nuclear fusion. Such computations were not possible in the middle of 1980's and 1990's. This opens a new avenue for highly acurate 
investigation of the bound state properties of the weakly-bound (1,1)-states in the both $(d d \mu)^{+}$ and $(d t \mu)^{+}$ ions. Moreover, currently we can predict probabilities of 
all essential processes in the five- and six-body $a b \mu e_2$ and $a b c \mu e_2$ muonic clusters. This allows us to describe (theoretically) all important processes in such systems 
to a very good accuracy. In turn, this gives us a hope for a resurgence of interest to the resonance muon-catalized fusion of nuclear reactions.

\newpage
\newpage
\begin{table}[tbp]
   \caption{The total energies of the (1,1)-states (or $P^{*}(L = 1)-$states) of the $(d d \mu)^{+}$ and $(d t \mu)^{+}$ muonic molecular 
            ions in muon-atomic units ($m.a.u.$) The notation $N$ is the total number of basis functions, Eq.(\ref{exp1}), used. In these 
            calculations we used the `old', `new' and `recent' sets of particle masses.}  
     \begin{center}
     \scalebox{0.95}{%
     \begin{tabular}{| c | c | c | c |}
      \hline\hline
 $N$  & $(d t \mu)^{+}$ (`old' set of masses) & $(d t \mu)^{+}$ (`new' set of masses) & $(d t \mu)^{+}$ (`recent' set of masses) \\  
     \hline
 3400 & -0.481991 529973 5424 & -0.481991 527054 3153 & -0.481991 526589 2129 \\
 3600 & -0.481991 529973 6396 & -0.481991 527054 4132 & -0.481991 526589 3102 \\ 
 3800 & -0.481991 529973 6974 & -0.481991 527054 4471 & -0.481991 526589 3679 \\  
 4000 & -0.481991 529973 7963 & -0.481991 527054 5698 & -0.481991 526589 4668 \\  
 4200 & -0.481991 529973 8646 & -0.481991 527054 6381 & -0.481991 526589 5351 \\ 
 4400 & -0.481991 529973 8919 & -0.481991 527054 6654 & -0.481991 526589 5624 \\
 4600 & -0.481991 529973 9021 & -0.481991 527054 6757 & -0.481991 526589 5727 \\
         \hline \hline
 $N$  & $(d d \mu)^{+}$ (`old' set of masses) & $(d d \mu)^{+}$ (`new' set of masses) & $(d d \mu)^{+}$ (`recent' set of masses) \\  
     \hline
 3400 & -0.473686 733842 725887 & -0.473686 731121 137705 & -0.473686 729773 104557 \\
 3600 & -0.473686 733842 726035 & -0.473686 731121 137853 & -0.473686 729773 104705 \\
 3800 & -0.473686 733842 726219 & -0.473686 731121 138038 & -0.473686 729773 104890 \\
 4000 & -0.473686 733842 726265 & -0.473686 731121 138085 & -0.473686 729773 104938 \\
    \hline \hline
  \end{tabular}}
  \end{center}
  \end{table} 
\end{document}